\documentclass[12pt]{article}
\usepackage{epsfig}

\begin{document}

\title{The temperature dependent performance analysis of EDFAs pumped at 1480 nm:
A more accurate propagation equation}
\author{C\"uneyt Berkdemir$\thanks{Corresponding author:
berkdemir@erciyes.edu.tr}$ {\small( OSA Member)}~~and~~Sedat \"Ozsoy\\
{\small {\sl Department of Physics, Faculty of Arts and Sciences,
Erciyes University,}} {\small {\sl 38039, Kayseri, Turkey}}}

\maketitle

\begin{abstract}
An analytically expression for the temperature dependence of the
signal gain of an erbium-doped fiber amplifier (EDFA) pumped at
1480 \textit{nm} are theoretically obtained by solving the
propagation equations with the amplified spontaneous emission
(ASE). It is seen that the temperature dependence of the gain
strongly depends on the distribution of population of
Er$^{3+}$-ions in the second level. In addition, the output pump
power and the intrinsic saturation power of the signal beam are
obtained as a function of the temperature. Numerical calculations
are carried out for the temperature range from $-$ 20 to $+$ 60
\textit {$^{o}$C} and the various fiber lengths. But the other
gain parameters, such as the pump and signal wavelengths and their
powers, are taken as constants. It is shown that the gain
decreases with increasing temperature within the range of $L\leq
27~m$.
\end{abstract}

\baselineskip=22pt plus 1pt minus 1pt
\hspace{-0.3cm}
\vspace{-0.3cm}

links and subsystems.

\section{Introduction}
EDFAs have been presenting many advantages such as high gain and
low noise in the optical communication networks and providing a
broadband amplification of radiation whose wavelength is in the
so-called third window for fiber-optic communication ($\sim$1530
\textit{nm}). In addition, the temperature dependence of the gain
characteristics of EDFAs has also of great importance for WDM
systems \cite{kem}. An analytical solution of the rate equation
has been derived and the gains at optimum amplifier lengths have
been experimentally found for the various temperature values in
the previous works \cite{per, nob, wei, furusawa}. The theoretical
and experimental results for the temperature dependence of the
gain with the various lengths of EDFAs have been reported
\cite{yam}, but the temperature dependent analytic expressions has
not been given in terms of the Boltzmann factor. Afterwards, the
theoretical analysis of amplification characteristics of EDFAs has
been developed to calculate the signal gain, using the rate
equation model, \cite{mao}, and this rate equation has been
modified by including the temperature and cross section factors to
understand the dependence of the gain on the temperature
\cite{fran, cun} for EDFAs pumped at 1480 \textit{nm}. In this
article, we present an analytical expression for the signal gain
in EDFAs, using the propagation equations improved by including
the temperature effects, and the numerical results for the
temperature ranges of $-$ 20 \textit{$^{o}$C} to $+$60 \textit
{$^{o}$C}. We took into account the amplified spontaneous emission
(ASE), but neglected the excited state absorption (ESA) effect for
the simplicity.

\section{Theory}
The simplest treatment of EDFA considers the two-level
amplification system with energy levels as shown in Figure 1, when
it is pumped at 1480 \textit{nm}. In this figure, level 1 is the
ground level and level 2 is the metastable level characterized by
a long lifetime $\tau$ ($=\gamma^{-1}$), $R_p^{a, e}$ is the pump
absorption and stimulated emission rates, $S_{12, 21}$ is the
signal stimulated absorption and emission rates, respectively.
$N_{2+}$ and $N_{2-}$ are the populations of $Er^{3+}$ ions within
the sub-levels of the second energy state and it is possible to
consider each of them as a single energy level. Actually, this
system contains many sub-levels where the erbium-ions reside and
they are unequally populated due to the thermal distribution of
the ions. Thus, the relative occupation of the sub-levels in the
thermal equilibrium must be arranged as a function of the
temperature. This arrangement is governed by Boltzmann's
distribution law:
\begin{equation} \label{1}
\beta=\frac{N_{2+}}{N_{2-}}=\frac{C_{nr}^+}{C_{nr}^-}=
exp\left(-{\frac{\Delta{E_{2}}}{k_{B}T}}\right)
\end{equation}
where \textit{T} is the temperature in degrees Kelvin,
\textit{$k_{B}$} is Boltzmann's constant. $E_{2+}$ and $E_{2-}$
are the higher and lower sub-levels energies of the second level,
respectively, and $\Delta{E_{2}= E_{2+}-E_{2-}}$ \cite{desurvire}.
$C_{nr}^+ $ and $C_{nr}^-$ are the nonradiative rates which
correspond to the thermalization process occurring within each
manifold of the second level. The rate equations corresponding to
the two levels 1 and 2 can be given as follow
\begin{equation} \label{2}
\frac{dN_{2+}}{dt}=R_p^aN_1-R_p^eN_{2+}+C_{nr}^-N_{2-}-C_{nr}^+N_{2+},
\end{equation}
\begin{equation} \label{3}
\frac{dN_{2-}}{dt}=S_{12}N_1-S_{21}N_{2-}-N_{2-}\gamma
-C_{nr}^-N_{2-}+C_{nr}^+N_{2+},
\end{equation}
\begin{equation} \label{4}
\frac{dN_{1}}{dt}=R_p^eN_{2+}-R_p^aN_1+S_{21}N_{2-}-S_{12}N_1+N_{2-}\gamma.
\end{equation}
Thus, at stationary conditions we now obtain
\begin{equation} \label{5}
N_{2-}=\tau\left[(\sigma_{p}^{a}N_{1}-\beta\sigma_{p}^{e}N_{2-})\frac{I_{p}}{h\nu_{p}}+(\sigma_{s}^{a}N_{1}-\sigma_{s}^{e}N_{2-})\frac{(I_{s}+I_{ASE}^{\pm})}{h\nu_{s}}\right],
\end{equation}
or
\begin{equation} \label{6}
\frac{N_{2-}}{N}=\frac{\displaystyle\frac{I_{p}}{b_{p}^{a}}+\frac{(I_{s}+I_{ASE}^{\pm})}{b_{s}^{a}}}{\displaystyle(1+\beta)\frac{I_{p}}{b_{p}^{a}}+
\beta\frac{I_{p}}{b_{p}^{e}}+(1+\beta+\eta)\frac{(I_{s}+I_{ASE}^{\pm})}{b_{s}^{a}}+1}
\end{equation}
where the populations are time invariant, i.e., $dN_i/dt=0$ ($i=1,
2$). In the last two equations,
$b_p^{a,e}=h\nu_p/\tau\sigma_p^{a,e}$,
$b_s^{a,e}=h\nu_s/\tau\sigma_s^{a,e}$, $\nu_p$ and $\nu_s$ are the
pump and the signal frequencies, respectively; $\sigma_p^{a,e}$ is
the stimulated absorption and emission cross sections of the pump
beam while $\sigma_s^{a,e}$ is the the stimulated absorption and
emission cross sections of the signal beam, respectively; $I_p$
and $I_s$ are the pump and signal intensities and $I_{ASE}^{\pm}$
is the forward (+ sign) and backward (- sign) propagating optical
intensities, respectively. $\eta$ is the ratio between the signal
emission and absorption cross sections, and the total
concentration distribution of $Er^{3+}$ ions is $N$,
$N=N_1+N_{2-}+N_{2+}$ or in terms of $\beta$,
$N=N_1+(1+\beta)N_{2-}$.

The differential equations for propagation of the signal, pump and
ASE powers are given, respectively, as follows
\begin{equation} \label{7}
\frac{dP_{s}}{dz}=2\pi\int_{0}^{\infty}I_{s}\left[\sigma_{s}^{e}N_{2-}(r)-\sigma_{s}^{a}N_{1}(r)\right]rdr,
\end{equation}
\begin{equation} \label{8}
\frac{dP_{p}}{dz}=\pm
2\pi\int_{0}^{\infty}I_{p}\left[\beta\sigma_{p}^{e}N_{2-}(r)-\sigma_{p}^{a}N_{1}(r)\right]rdr,
\end{equation}
\begin{equation} \label{9}
\frac{dP_{ASE}^{\pm}}{dz}=\pm
2h\nu_s\int_{0}^{\infty}2\pi\sigma_{s}^{e}N_{2-}f_{ASE}^{\pm}(r)rdr
\pm2\pi\int_{0}^{\infty}\left[\sigma_{s}^{e}N_{2-}(r)-\sigma_{s}^{a}N_{1}(r)\right]P_{ASE}^{\pm}f_{ASE}^{\pm}rdr,\\
\end{equation}
where $f_{ASE}^{\pm}$ is the normalized ASE intensity profile,
$P_{ASE}^{\pm}$ is the amplified spontaneous emission power at the
position z and has to be determined from a forward as well as a
backward travelling ASE spectrum,
\begin{equation} \label{10}
P_{ASE}^{\pm}=P_{ASE}^{+}+P_{ASE}^{-}.
\end{equation}
We can decompose the intensity as $I_{s,p}(z,
r)=P_{s,p}(z)f_{s,p}(r)$ where $P_{s,p}(z)$ is z-dependent signal
od pump powers and $f_{s,p}(r)$ is the normalized signal and pump
transverse intensity profiles, respectively. At this point, by
substituting $N=N_1+(1+\beta)N_{2-}$ into Eq.(\ref{7}), we have
the propagation equation for the signal power:
\begin{equation} \label{11}
\frac{dP_{s}}{dz}=2\pi\sigma_{s}^{a}P_{s}(1+\beta
+\eta)\int_{0}^{\infty}N_{2-}f_s(r)rdr-P_{s}\alpha_s,
\end{equation}
where $\alpha_s=2\pi\sigma_{s}^{a}\int_{0}^{\infty}N(r)f(r)rdr$ is
the absorption constant of the signal beam. To evaluate the
integral at the right-hand side of Eq.(\ref{11}), we make use of
Eq.(\ref{5}). In this case, multiplying both-hand side of
Eq.(\ref{5}) with $rdr$ and then integrating between $0$ and
$\infty$, we obtain the following equations:
\begin{eqnarray} \label{12}
\int_{0}^{\infty}N_{2-}rdr=\int_{0}^{\infty}\frac{\tau
I_{p}}{h\nu_{p}}(\sigma_{p}^{a}N_{1}-\beta\sigma_{p}^{e}N_{2-})rdr
+\int_{0}^{\infty}\frac{\tau I_{s}}{h\nu_{s}}(\sigma_{s}^{a}N_{1}-\sigma_{s}^{e}N_{2-})rdr\nonumber\\
+\int_{0}^{\infty}\frac{\tau
I_{ASE}^{+}}{h\nu_{s}}(\sigma_{s}^{a}N_{1}-\sigma_{s}^{e}N_{2-})rdr,
\end{eqnarray}
\begin{eqnarray} \label{13}
\int_{0}^{\infty}N_{2-}rdr=-\frac{\tau}{2\pi
h\nu_{p}}\frac{dP_p}{dz} -\frac{\tau}{2\pi
h\nu_{s}}\frac{dP_s}{dz}- \frac{\tau}{2\pi
h\nu_{s}}\frac{dP_{ASE}^+}{dz}+2\tau\sigma_{s}^{e}\int_{0}^{\infty}N_{2-}f(r)rdr,
\end{eqnarray}
\begin{eqnarray} \label{14}
\int_{0}^{\infty}N_{2-}f(r)rdr=-\frac{\tau}{2\pi
(A/\Gamma-2\tau\sigma_{s}^{e})}\left[\frac{1}{h\nu_{p}}\frac{dP_p}{dz}
+\frac{1}{h\nu_{s}}\left(\frac{dP_s}{dz}+\frac{dP_{ASE}^+}{dz}\right)\right],
\end{eqnarray}\\
where we define the confinement factor $\Gamma
=A\int_{0}^{\infty}N_{2-}f(r)rdr/\int_{0}^{\infty}N_{2-}rdr$ and A
is the effective doped area. We can put the equations into more
practical form supposing the pump, signal and ASE profiles to be
approximately equal, so that the transverse profiles $f_p(r)\sim
f_s(r)\sim f_{ASE}^+(r)=f(r)$ and considering the co-propagating
scheme in the positive z direction for the simplicity. Inserting
Eq.(\ref{14}) into Eq.(\ref{11}), we have
\begin{eqnarray} \label{15}
\frac{dP_{s}}{dz}=-P_{s}\left(\alpha_s+\frac{h\nu_{s}}{P_{s}^{int}}\left[\frac{1}{h\nu_{p}}\frac{dP_p}{dz}
+\frac{1}{h\nu_{s}}\left(\frac{dP_s}{dz}+\frac{dP_{ASE}^+}{dz}\right)\right]\right),
\end{eqnarray}
where the intrinsic saturation power of the signal beam is
introduced as follow
\begin{equation} \label{16}
P_{s}^{int}=h\nu_{s}(A-2\tau\sigma_{s}^{e}\Gamma)/\tau\sigma_{s}^{a}\Gamma(1+\beta
+\eta).
\end{equation}
Therefore, we define the intrinsic saturation power as a function
of the temperature. Integrating Eq.(\ref{15}), we obtain the
output signal power at $z=L$ and hence establish a relationship
between the amplifier gain and length:
\begin{equation} \label{17}
\frac{P_{s}(L)}{P_{s}(0)}=exp(-\alpha_sL)
exp\left(\frac{h\nu_{s}}{P_{s}^{int}}\left[\frac{P_p(0)-P_p(L)}{h\nu_{p}}+\frac{(P_s(0)+P_{ASE}^{+}(0))-(P_s(L)
+P_{ASE}^{+}(L))}{h\nu_{s}}\right]\right).
\end{equation}
The amplifier gain $G=P_{s}(L)/P_{s}(0)$ can be calculated from
the following equation:
\begin{equation} \label{18}
G=exp(-\alpha_sL)
exp\left(\frac{h\nu_{s}}{P_{s}^{int}}\left[\frac{P_p(0)-P_p(L)}{h\nu_{p}}-\frac{P_s(0)}{h\nu_{s}}(1-G)
-\frac{P_{ASE}^{+}(L)}{h\nu_{s}}\right]\right),
\end{equation}
with boundary condition
\begin{equation} \label{19}
P_{ASE}^{+}(0)=0.
\end{equation}
If one neglects the effect of the $\beta$ parameter and the ASE
power in the gain equation, it can be easily seen that the
relevant equation is reduced to the previous works \cite {per,
lin}. Thus, Eq.(\ref{19}) is a more accurate solution for the
propagation equations. In order to obtain the output pump power
$P_p(L)$ in Eq.(\ref{18}) for the maximal pumping efficiency, it
should be substituted Eq.(\ref{6}) into Eq.(\ref{7}) and
Eq.(\ref{8}), and then Eq.(\ref{7}) divided by Eq.(\ref{8}). If
the obtained result makes equal to zero, we have
\begin{equation} \label{20}
P_p(L)=\frac{1}{R\left(\frac{\eta}{b_p^a}-\frac{\beta}{b_p^e}\right)},
\end{equation}
where $R=\int_{0}^{\infty}N(r)f(r)rdr/\int_{0}^{\infty}N(r)rdr$.
It is notes that the output pump power is a function of the
temperature.

\section{Results and discussion}
The gain against fiber lengths is calculated in the following way.
Firstly, we take $f(r)$ in Gaussian form,
$f(r)=exp(-r^2/\omega_0^2)/\pi\omega_0^2$ where $\omega_0$ is the
spot size and the effective core area is $\pi\omega_0^2=33~\mu
m^2$. Dopant distribution $N(r)$ is also assumed to be Gaussian,
$N(r)\simeq exp(-r^2/\omega^2)/\pi\omega^2$. In addition, the
ratio ($\omega /\omega_0$) between Gaussian dopant distribution
and transverse intensity profiles is selected as $0.3$. Secondly,
we obtain $R$ and $\alpha_s$ by using $N(r)$ and $f(r)$ for the
relevant fiber parameters. Thus, the output pump power in
Eq.(\ref{20}) is calculated with the different temperature values
for the fiber length of $45~m$. In this case, it is bear in mind
that the ratio of cross-sections, which are belong to the signal
beam, depends on the temperature. To calculate the parameter
$\eta$ as a function of the temperature, we benefit by McCumber's
theory, which gives a highly accurate relation between emission
and absorption cross sections \cite {zech}.

In the numerical calculations, we select the Al/P-silica
erbium-doped fiber as an amplifier operated at the pump wavelength
$\lambda_p=1480~nm$ and the input pump power $P_p(0)$ is fixed at
30 $mW$. The signal wavelength $\lambda_s$ and the signal power
$P_s(0)$ are taken as $1530~nm$ and 10 $\mu W$, respectively. The
other parameters assigned to the fiber are given in Table 1 \cite
{lin}. Moreover, we used the simulation programme {\it
OptiAmplifier 4.0} for generating $P_{ASE}^{+}(L)$ only, and we
set up the basic system seen in Figure 2 \cite {opti}. The energy
difference between the sublevels of the metastable level (level 2)
is assumed as $300~cm^{-1}$ in the room temperature for the
simplicity.

The results calculated for the various output pump powers and the
intrinsic saturation powers as well as the parameters $\beta$ and
$\eta$ are given in Table 2. The variation of the signal gain
against the fiber length is illustrated in Figure 3 for the
temperatures $-20~^oC$, $20~^oC$ and $60~^oC$. For a given pump
and signal powers the gain decreases with increasing temperature
within the range of $L\leq 27~m$. The difference between the
maximum gains for $-20$ and $60~^oC$ is $0.67~dB$. There is a
temperature insensitivity for the length about $L\approx 30~m$ for
the relevant pump and signal powers. On the other hand, this
temperature insensitive length is equivalent to the length at
which the gain curves intersect each other (Figure 3).

\section{Conclusion}
We have introduced a more accurate model including the temperature
effect for the signal gain of the erbium-doped fiber amplifier. In
addition, we have shown the possibility of deriving an analytical
solution of the propagation equations for some practical
temperature ranges. The temperature dependence of the output pump
power is smaller than that of the intrinsic saturation power.
Thus, in terms of the practical applications we can neglect the
dependence of the output pump power on the temperature. However,
it is taken into consideration that the gain performance of EDFAs
strongly depends on the temperature.\\

\noindent {\bf Acknowledgements}

This study is supported by Scientific Research Projects Council
(SRPC) of Erciyes University under Grant FBT-04-17. The authors
are grateful to A. ALTUNCU for his useful comments and discussions
on the original version of the paper.

\newpage

{\bf Table 1:}~~{\small Typical fiber parameters for an Al/P-silica erbium-doped fiber (from Ref.\cite{lin}).\\

\begin{tabular}{ccc}\hline\hline
\\ $\mathbf {Symbols}
\hspace*{0.2cm} $ & $ \hspace*{0.2cm} \mathbf{Definations}
\hspace*{0.2cm} $ & $ \hspace*{0.2cm} \mathbf{Values} $ \\[0.3cm]\hline
~~~~&~~~~~~~~&~~\\[0.2cm]
${\sigma_s^e}$&~~signal~emission~cross-section~~~&~~~~~~$5.7x10^{-25}$~$m^2$~ \\[0.2cm]
$\sigma_s^a$&~~signal~absorption~cross-section~~~&~~~~~~$6.6x10^{-25}$~$m^2$~ \\[0.2cm]
$\sigma_p^e$&~~pump~emission~cross-section~~~&~~~~~~$0.87x10^{-25}$~$m^2$~ \\[0.2cm]
$\sigma_p^a$&~~pump~absorption~cross-section~~~&~~~~~~$2.44x10^{-25}$~$m^2$~ \\[0.2cm]
$\tau$&~~life~time~&~~~~~~10.8~$ms$~~ \\[0.2cm]
$N$&~~erbium~concentration~&~~~~~~~~$3.86x10^{24}$~$m^{-3}$~~ \\[0.2cm]
$\lambda_s$&~~signal~wavelength~&~~~~~~1530~$nm$~~ \\[0.2cm]
$\lambda_p$&~~pump~wavelength~&~~~~~~1480~$nm$~~ \\[0.2cm]
$P_{ASE}^+(L)$&~~copropagating~ASE~power&~~~~~~0.15~$mW$~~ \\[0.2cm]\hline
~~~~&~~~~&~~~~~~ \\[0.2cm]
\end{tabular}

\newpage

{\bf Table 2:}~~{\small The relevant fiber parameters as a function of temperature.\\

\begin{tabular}{cccccc}\hline\hline\\
$\mathbf {Temperature~(^oC)} \hspace*{0.2cm} $ & $\hspace*{0.2cm}
\mathbf{\beta} \hspace*{0.2cm} $ & $\hspace*{0.2cm} \mathbf{\eta}
\hspace*{0.2cm} $ & $ \hspace*{0.2cm} \mathbf{P_p(L)}
\hspace*{0.2cm} $ & $ \hspace*{0.2cm} \mathbf{P_s^{int}} $ \\[0.3cm]\hline
~~&~~~~&~~~&~~~&~~\\[0.2cm]
$-~20$&~0.306~~&~~0.845~~&~~2.308~$mW$&~~0.493~$mW$~~ \\[0.2cm]
$+~20$&~0.357~~&~~0.862~&~~2.311~$mW$&~~0.474~$mW$~~ \\[0.2cm]
$+~60$&~0.406~~&~~0.879~&~~2.314~$mW$&~~0.451~$mW$~~~ \\[0.2cm]\hline
~~~~&~~~~&~~&~~~&~~~ \\[0.2cm]
\end{tabular}

\newpage

\vspace{0.5in} \noindent{\Large \bf Figure Captions} \vskip .5
true cm

\noindent {\bf Figure 1:} Two level amplification system and main transition of erbium ion.\\

\noindent {\bf Figure 2:} Simulation setup for measurement of the
co-propagating ASE power in an Er$^{3+}$-doped optical fiber amplifier (from {\it OptiAmplifier 4.0}).\\

\noindent {\bf Figure 3:} Gain as a function of fiber length. $P_p(0)=30~mW$ and $P_s(0)=10~\mu W$.\\

\newpage






\end{document}